\documentclass[twocolumn,times]{aastex}

\usepackage{graphicx}
\usepackage{natbib}
\usepackage{color}
\usepackage{tabularx}
\usepackage{bm}

\DeclareGraphicsExtensions{.pdf,.png,.jpg}

\newcommand{\besancon}{Besan{\c c}on }

\newcommand{\kepler}{\emph{Kepler}}

\newcommand{\mabuls}{{\sc{MaB$\mu$LS}}}

\newcommand{\thetae}{\theta_{\mathrm{E}}}

\newcommand{\pie}{\pi_{\mathrm{E}}}
\newcommand{\pien}{\pi_{\mathrm{E,N}}}
\newcommand{\piee}{\pi_{\mathrm{E,E}}}
\newcommand{\pirel}{\pi_{\mathrm{rel}}}
\newcommand{\vecpie}{\bm{\pi}_{\mathrm{E}}}
\newcommand{\dl}{D_{\mathrm{l}}}
\newcommand{\ds}{D_{\mathrm{s}}}

\newcommand{\tein}{t_{\mathrm{E}}}
\newcommand{\murel}{\mu_{\mathrm{rel}}}
\newcommand{\tzero}{t_{\mathrm{0}}}

\newcommand{\uzero}{u_{\mathrm{0}}}

\newcommand{\zone}{z_{\mathrm{1}}}

\newcommand{\msun}{M_{\odot}}

\newcommand{\mjup}{M_{\mathrm{Jupiter}}}

\newcommand{\percent}{\%}

\newcommand{\nhigh}{7.9}
\newcommand{\nlow}{1.4}
\newcommand{\noglehigh}{5.1}

\newcommand{\npiehigh}{3.9}
\newcommand{\npielow}{1}

\newcommand{\nrholow}{0.4}
\newcommand{\nmasshigh}{0.98}
\newcommand{\nmasslow}{0.42}
\newcommand{\fchihigh}{17}
\newcommand{\fchilow}{30}

\newcommand{\probnohigh}{0.00038}
\newcommand{\probnolow}{0.24}
\newcommand{\chiratio}{77}
\newcommand{\medIhigh}{20.4}
\newcommand{\medIlow}{19.1}
\newcommand{\medVhigh}{22.5}
\newcommand{\medVlow}{21.0}

\newcommand{\kt}{{\it K2}}
\newcommand{\ktcn}{{\it K2}C9}
\newcommand{\spitzer}{{\it Spitzer}}

\newcommand{\gulls}{{\sc gulls}}

\newcommand{\tzkep}{t_{0,Kep}}
\newcommand{\uzkep}{u_{0,Kep}}
\newcommand{\tekep}{t_{{\rm E},Kep}}
\newcommand{\vecmurel}{\bm{\mu}_{\rm rel}}
\newcommand{\reproj}{\tilde{r}_{\rm E}}
\newcommand{\dbase}{D_{\perp}}

\shorttitle{{\it K2}C9 Free-Floating Planet Predictions}
\shortauthors{Penny et al.}
\begin{document}

\title{Predictions for the Detection and Characterization of a Population of Free-Floating Planets with {\it K2} Campaign 9}

\author{Matthew T. Penny,\altaffilmark{1,2} Nicolas J. Rattenbury,\altaffilmark{3} B. Scott Gaudi\altaffilmark{1} and Eamonn Kerins\altaffilmark{4}}
\email{penny@astronomy.ohio-state.edu}

\altaffiltext{1}{Department of Astronomy, Ohio State University, 140 West 18th Avenue, Columbus, OH 43210, USA}
\altaffiltext{2}{Sagan Fellow}
\altaffiltext{3}{Department of Physics, University of Auckland, Private Bag 92019, Auckland, New Zealand}
\altaffiltext{4}{Jodrell Bank Centre for Astrophysics, School of Physics and Astronomy, University of Manchester, Oxford Road, Manchester M13 9PL, UK}

\slugcomment{Submitted to AAS Journals}

\begin{abstract}
\kt\ Campaign 9 (\ktcn) offers the first chance to measure parallaxes and masses of members of the large population of free-floating planets (FFPs) that has previously been inferred from measurements of the rate of short-timescale microlensing events.
Using detailed simulations of the nominal campaign (ignoring the loss of events due to \kepler's emergency mode) and ground-based microlensing surveys, we predict the number of events that can be detected if there is a population of $1$-$\mjup$ FFPs matching current observational constraints.
Using a Fisher matrix analysis we also estimate the number of detections for which it will be possible to measure the microlensing parallax, angular Einstein radius and FFP mass.
We predict that between $\nlow$ and $\nhigh$ events will be detected in the \kt\ data, depending on the noise floor that can be reached, but with the optimistic scenario being more likely.
For nearly all of these it will be possible to either measure the parallax or constrain it to be probabilistically consistent with only planetary-mass lenses.
We expect that for between $\nmasslow$ and $\nmasshigh$ events it will be possible to gain a complete solution and measure the FFP mass.
For the emergency mode truncated campaign, these numbers are reduced by $20$~percent.
We argue that when combined with prompt high-resolution imaging of a larger sample of short-timescale events, \ktcn\ will conclusively determine if the putative FFP population is indeed both planetary and free-floating.
\end{abstract}

\section{Introduction}\label{intro}

The large population of free-floating or loosely-bound, Jupiter-mass planets (hereafter FFPs) inferred by \citet{Sumi2011} remains difficult to explain. 
After accounting for various possible forms of the stellar and sub-stellar mass function, \citet{Sumi2011} found that an excess of short-timescale microlensing events could be explained by a population of $1.9_{-0.8}^{+1.3}$ Jupiter-mass objects per main sequence star (when assuming a three-component power-law stellar mass fuction). 
Extrapolation of the measured low-mass IMFs of nearby, young clusters and associations~\citep[e.g.][]{PenaRamirez2012,Lodieu2013}, beyond that done by \citet{Sumi2011} and down to 1~Jupiter mass can only explain about ${\sim}10$~percent of the events that are seen. 
Statistical surveys of wide-separtion, young exoplanets with high-contrast imaging~\citep{Quanz2012, Bowler2015,Reggiani2016,Durkan2016} place strong limits on the abundance of bound planets that could potentially be mistaken for FFPs, even with cold-start models where planets are expected to be faint. 
Finally, synthesizing the results of surveys that cover a wide range of parameter space~\citep{Clanton2014a,Clanton2016} and summing up all of the planetary mass and assuming upper limits are measurements yields a result that is at least a factor of two smaller than the amount of mass locked up in the inferred FFP population~\citep{Henderson2015}.

Theory does not seem to offer a convenient way out of the impasse.
\citet{Veras2012a} found that tens of giant planets would need to be formed by each star in order to explain the measured FFP abundance if planet-planet scattering were the cause, and simulations by \citet{Pfyffer2015} demonstrate that more modest planetary systems cannot eject a sufficient number of giant planets.
Without a central, stationary gravitational sink, circumbinary planetary systems have recently been shown to eject significantly more giant planets than single-star systems~\citep{Sutherland2016,Smullen2016}, but only a small fraction of stellar systems are binaries with sufficiently small orbits to form circumbinary systems~\citep[${\sim}$10~percent with periods less than 1000~d][]{Raghavan2010}, and so giant planet formation and ejection in such systems would need to be unreasonably prolific.
A similar issue faces the channel of ejection during post main sequence evolution~\citep{Veras2016}.

\kt\ Campaign 9~\citep[\ktcn,][]{Howell2014,Henderson2015,Gould2013-whitepaper} offers the first opportunity to observationally challenge, or indeed confirm, the inference of \citet{Sumi2011} with more than just improved statistics.
Microlensing satellite parallax observations~\citep{Refsdal1966,Gould1994} enable the measurement of the microlens parallax $\vecpie$, which can place a strong constraint on the nature of the lens and go some way toward breaking the fundamental microlensing timescale degeneracy between mass, distances and relative velocity.
Satellite parallax measurements have now been made for a large number of events with \spitzer~\citep{Dong2007,Udalski2015-spitzer,Yee2015,CalchiNovati2015,Zhu2015-binary,Zhu2015-isolated,Shvartzvald2015-spitzer,Poleski2015-spitzer,Bozza2016,Street2016}, but the time required to identify events and arrange observations with \spitzer\ precludes its application toward FFPs.
Instead, one must blindly survey a large area of sky from Earth and a satellite simultaneously in order to have a chance to detect FFPs.
This is what \ktcn\ enables.

Here we perform simulations to determine if \ktcn\ can detect a sufficient number of FFP events to place interesting constraints on the FFP population.
In Section~\ref{sims} we describe our simulations.
In Section~\ref{results} we present our results, and in Section~\ref{discuss} we discuss their implications for characterizing the FFP population.
We conclude in Section~\ref{conclusions}.

\section{Simulations}\label{sims}

To simulate the combination of ground-based and \ktcn\ surveys, we used the simulation code originally presented by \citet{Penny2013}, which has been renamed {\sc gulls}.\footnote{The original name \mabuls\ now only refers to a tool for computing microlensing optical depth and event rates using the \besancon\ model~\citep{Awiphan2016}.} By producing artificial images, the code simulates photometry of gravitational microlensing lightcurves with lens and source stars drawn from a Galactic population synthesis model.
The code has been modified extensively to allow for the simulation of ground-based observations, to include the effects of parallax, and to add the estimation of event-by-event parameter uncertainties using Fisher matrix analysis.
As the general mechanics of the code are described in detail by \citet{Penny2013}, we only describe the modifications here.
There are a few instances where mistakes were made in setting up the simulations, but were not significant enough to justify rerunning the simulation; in each case we note the mistake below.

\subsection{Lightcurves and their Parametrization}\label{parametrization}

In this paper we are only interested in FFP events, which are isolated lenses.
We therefore simulate single-lens lightcurves with finite-source effects but no limb-darkening,\footnote{We have previously experimented with incorporating limb darkening for bound Earth-mass exoplanet events, where it is likely to play a larger role than here, and found that it did not significantly affect detection efficiencies.} using the method of \citet{Witt1994} to compute the magnification.
We parameterize the lightcurve using the time of closest approach between the lens and source as seen by the \kepler\ spacecraft $\tzkep$, the impact parameter of this closest approach $\uzkep$, and the Einstein radius crossing time $\tekep$ in the inertial frame moving with \kepler\ at time $\tzkep$.
The relative angular size of the  source is described by $\rho=\theta_{\ast}/\thetae$, where $\theta_{\ast}$ is the angular radius of the source star and 
\begin{equation}
\thetae = \sqrt{\kappa M \pirel},
\label{thetae}
\end{equation}
is the angular Einstein radius, where $\kappa=8.144$~mas~$\msun^{-1}$ is a constant, $M$ is the lens mass and $\pirel=$AU$(\dl^{-1}-\ds^{-1})$ is the relative parallax, with $\dl$ and $\ds$ being the lens and source distance, respectively~\citep[see, e.g.,][]{Gould2000}.
Microlensing parallax is parametrized by the vector $\vecpie = (\pien,\piee)$, with components in the north and east directions, and is defined as
\begin{equation}
\vecpie = \frac{\pirel}{\thetae}\frac{\vecmurel}{\murel},
\end{equation}
where $\vecmurel$ is the vector relative lens-source proper motion measured in a heliocentric frame.
The impact parameter $\uzero$ and peak time $\tzero$ of the event seen from Earth depends on $\vecpie$ and the vector separation between the Earth and Kepler projected on the plane normal to the event's location on the sky.
$(\tzero,\uzero)$ and $(\tzkep,\uzkep)$ are related by
\begin{equation}
\left(\frac{\Delta\tzero}{\tekep},\Delta\uzero\right) = \left(\frac{\tzero-\tzkep}{\tekep},\uzero-\uzkep\right) = \frac{\bm{D}_{\perp}}{\reproj},
\label{dtdu}
\end{equation}
where the components of $\bm{D}_{\perp}$ are defined to be in the direction parallel and perpendicular to $\vecpie$, and $\reproj=\text{AU}/\pie$ is the projected Einstein radius.
We parameterize the effects of source and blended light with a baseline magnitude, e.g., $I_0$, and the fraction of this flux that is contributed by the source $f_{\rm s}$, both of which are different for each observatory and each filter.

\subsection{Ground-based Observatories}\label{gbobs}

To schedule observations, \gulls\ reads in a schedule script that is repeated for the duration of the simulations.
For ground-based observations this is overridden and observations are not taken if the elevation of an event falls below an elevation limit (here optimistically set at 20\degr), or if the sun's elevation is (again, rather optimistically) above -6\degr.
The basic astronomical functions used for these computations were based on code by \citet{Ofek2014}.
Observations are also not taken if the weather is ``bad;'' we incorporate into the definition of bad weather anything that might halt observations.
A fixed weather calendar is computed in advance with a site-by-site good weather probability, that is tested against a uniform random deviate every six hours.
We removed by hand the nights on each calendar where the moon would have been too close to the survey fields for wide-field survey observations (the night of minimum separation, and one night either side of this). 

The architecture of our code makes it difficult to simulate the effect of variable seeing, so we simply opt to assume a single value of seeing, the median, for each site.
In each case we used a \citet{Moffat1969} PSF with parameter $\beta=4$.
We do, however, simulate the effect of variable sky brightness and extinction as a function of filter, elevation and moon position using the model of \citet{Krisciunas1991}.
We also account for the increase of atmospheric extinction with airmass through a filter-dependent extinction coefficient.
The adopted parameters for ground-based observatories are described in \ref{params}. 

\subsection{Parallax and Orbits}\label{parallax}

To compute the effects of parallax on the lightcurve, we begin by defining an inertial reference frame, which in this case we choose to be the frame moving with \kepler\ at the time of the event's peak magnification as seen by \kepler\ $\tzkep$.
We compute the orbits of observers using the formalism and ephemerides of \citet{Standish2013}.
We approximate \kepler's orbit by assuming its orbital elements to be the same as the Earth's, except for the semimajor axis $a_0=1.01319$~AU, mean longitude $L_0=164.72296\degr$ and its rate of change $\dot{L}=352.99329\degr$~yr$^{-1}$, which were chosen to place \kepler\ at the same position as the Earth on its launch date and to match its orbital period of $372.5$~d.
The accuracy of the orbit is more than sufficent for our purposes in simulating the mission, and reproduces the projected separation between Earth and \kepler\ as a function of time, which is shown in Figure~\ref{t0dist} below.
Ground-based observers were assumed to be at the center of the Earth as it orbits the Earth-Moon barycenter. 

For each epoch of observation, we compute the position of the observer in the inertial reference frame and translate this into a 2-d vector shift in the apparent location of the source star, parallel and perpendicular to its linear trajectory in the inertial frame, in units of Einstein radii.
The shift depends on the sky coordinates of the microlensing event and the microlensing parallax.
A detailed description of satellite parallaxes in a heliocentric frame is given by \cite{CalchiNovati2015}.

\subsection{Fisher Matrix Parameter Estimates}\label{fisher}

For each simulated event, in addition to assessing whether the event will be detected or not, we would like to estimate what the measurement uncertainties of the event's parameters would be.
We do this using a Fisher matrix analysis~\citep[see, e.g.,][]{Gould2003}, by assuming that the input lightcurve model is the best fit to the data and that the distribution of $\chi^2$ is quadratic in each parameter.
Even when this assumption breaks down, a large value of a Fisher matrix parameter uncertainty estimate will indicate that a parameter is unconstrained by the data.
To compute the uncertainty on the magnitude of the parallax vector $\pie$ from the covariance matrix of the lightcurve parameters, we multiply the covariance matrix by the jacobian of the equation $\pie = \sqrt{\pien^2+\piee^2}$.

\subsection{Galactic Model}\label{galmod}

In contrast to \citet{Penny2013}, we use the public version of the \besancon\ model as our input Galactic model~\citep{Robin2003}.
We found it necessary to make the following changes to the star catalogs that are the output by the model's web interface.\footnote{http://model.obs-besancon.fr/}
First, we compute a larger range of stellar magnitudes than provided by the model, by converting unreddened MegaCam $ugriz$ magnitudes to SDSS $ugriz$ magnitudes using the transformations of \citet{Gwyn2008}.
Extinction was then applied to each magnitude using the \citet{Marshall2006} 3-d extinction model in the $K$-band.
$K$-band extinctions were converted to other bands by assuming a ratio of total to selective extinction $R_V=2.5$~\citep{Nataf2013} and a \citep{Cardelli1989} reddening law.
After applying extinction, we computed Johnson-Cousins magnitudes using the ``Lupton (2005)'' transformations\footnote{see https://www.sdss3.org/dr8/algorithms/sdssUBVRITransform.php}
\begin{eqnarray}
V =& g - 0.5784(g-r)-0.0038,\\
R =& r - 0.2936(r-i)-0.1439,\\
I =& i - 0.3780(i-z)-0.3974,
\end{eqnarray}
and \kepler\ magnitudes using the transformation from $g$ and $r$ magnitudes of \citet{Brown2011}.
In hindsight it would have been preferable to perform all the magnitude transformations before applying extinction, but the effect on the resultant magnitudes is tiny (e.g., for $V$, it results in a 0.3~percent change in the $V$-to-$K$ exctinction ratio $A_V/A_K$).

The second change corrects an error in the $V$-component of stellar $UVW$ velocities that affects stars beyond the position of the Galactic center ($X>8$~kpc in the standard heliocentric cartesian system).
The error was propogated to proper motions and so would affect our event rates if left uncorrected.
We applied the correction \citetext{A. Robin, private communication}
\begin{equation}
V \rightarrow -2 V_{\rm LSR} -V \quad\text{if}\quad X>8~\text{kpc},
\end{equation}
where $V_{\rm LSR}=226.4$~km~s$^{-1}$ is the local standard of rest, and recomputed proper motions~\citep{Johnson1987}.

In the \besancon\ model, each stellar population (thin disk, bulge, etc.) was allowed its own stellar IMF.
As the low-mass end of the bulge IMF was poorly constrained at the time~\citep{Robin2003}, it was chosen to be a continuous \citet{Salpeter1955} mass function with slope $-2.35$.
This over-produces low-mass stars in the bulge if the bulge IMF is similar to that of \citet{Kroupa2001} or \citet{Chabrier2003}, which while this barely affects star counts measurable from the ground, causes a microlensing event rate that is too high.
To correct for this, we weighted events involving bulge stars with masses below $0.5\msun$ by a factor $(M/0.5\msun)^1$, effectively giving the bulge an IMF similar to that of the other populations in the model, with a low-mass slope of $-1.35$.
The weighting is applied twice if both the source and lens stars have masses below $0.5\msun$.

Even after adjusting the mass function in the bulge, the number of bulge main sequence stars was still too high, as has been recognized by \citet{Kerins2009}.
Our final change to the model was to further weight events involving bulge stars by a factor $f$, equal to match the overprediction of main sequence bulge stars in the \besancon\ model to the number measured by \citep{Calamida2015} using HST data at $(\ell,b)=(1\fdg25,-2\fdg65)$.
This factor was $f=1/3.16$.
Again, if both source and lens stars belonged to the bulge, this down-weighting was applied twice.

\begin{deluxetable}{lcccc}
\tabletypesize{\footnotesize}
\tablecolumns{5}
\tablecaption{Simulation parameters.}
\label{simparams}
\tablehead{\colhead{Parameter} & \colhead{\kt} & \colhead{OGLE} & \colhead{OGLE} & \colhead{MOA}}
\startdata
Filter & $K_p$ & $V$ & $I$ & MOA-$R$ \\
Cadence (min) & 30 & 45.5 & 45.5 & 10 \\
Exposure time & $270\times6$~s & 100~s & 150~s & 60~s \\
Start Observing & Apr 6 & \multicolumn{3}{c}{full 2016 season} \\
End Observing & Jun 29 & --- & --- & --- \\
\hline
Pixel size (\arcsec) & 3.98 & 0.26 & 0.26 & 0.58 \\
PSF size & ---$^{\ast}$ & $0\farcs9$ & $0\farcs9$ & $2\farcs0$ \\
Gain (e$^-$/ADU) & 110 & 1.6 & 1.6 & 2.2 \\
Readout noise (e$^-$) & 120 & 7.5 & 7.5 & 6 \\
Full well ($10^5$ e$^-$) & 10 & 2.0 & 2.0 & 2.0 \\
Bits pixel$^{-1}$ & 14 & 16 & 16 & 16 \\
$m_{\rm zero}$ $^{\dag}$ & 12 & 22 & 22 & 22 \\
\vspace{-2pt}Flux at $m_{\rm zero}$ & $1.83\times10^5$ & 9.577 & 5.475 & 22.65 \\
(e$^-$~s$^{-1}$) & & & & \\
\vspace{-2pt}Systematic error$^{\ddag}$ & 0.1\percent\ (O) & 0.4\percent$^{\ast\ast}$ & 0.4\percent & 0.4\percent \\
 & 45~e$^-$~s$^{-1}$ (P) & & & \\
Aperture radius & $3\times3$~pixels & $0\farcs9$ & $0\farcs9$ & $2\farcs0$ \\
\vspace{-2pt}Sky$^{\dag\dag}$ & 21.5$^{\ddag\ddag}$ & 21.8 & 19.9 & 20.2 \\
(mag~arcsec$^{-2}$) & & & & \\
\vspace{-2pt}Extinction coefficient & --- & 0.14 & 0.069 & 0.075 \\
(mag~airmass$^{-1}$) & & & &\\
\vspace{-2pt}Good-weather & --- & 0.75 & 0.75 & 0.60 \\
probability & & & & \\
\enddata
\tablecomments{Simulation parameters for each of the observatories.\\ $^{\ast}$For the \kepler\ PSF we use the numerical PSF from detector 10 in module 4 at the edge of the focal plane~\citep{Bryson2010}, which should be very similar to the PSFs in the superstamp.\\ $^{\dag}$Magnitude at which flux per second is defined.\\ $^{\ddag}$For \kt\ we simulate two scenarios for a systematic noise floor, optimisitc (labeled O above) and pessimistic (labeled P).\\ $^{\ast\ast}$Adopted from \citet{Henderson2014-kmt}.\\ $^{\dag\dag}$Sky background at zenith with no moonlight.\\ $^{\ddag\ddag}$For all observers we include a model of the zodiacal light~\citep{Leinert1998}, but for \kepler\ we mistakenly double counted and added a constant background of the roughly the same brightness; the impact should be minimal however as stars will bring every pixel ``above sky.'' }
\end{deluxetable}

The above corrections were applied in turn, instead of simply computing a single overall event rate scaling factor, in order to make sure that the distance distribution of lenses would remain as realistic as possible.
After making the above corrections, we tested the event rates of the adjusted model by simulating in detail the MOA-II survey conducted in 2006 and 2007~\citep{Sumi2011, Sumi2013}, matching the weather patterns and detection cuts to predict the number of events that actually enter the sample.
A full description of this simulation will be presented in a future paper.
The simulation predicted (or more accurately postdicted) the survey would find a factor of 0.59 times the number of events that were actually found and passed all detection cuts.
The timescale distribution of the simulation was a good match to that of the data.
We therefore multiply the results of our simulations by $1/0.59$ to match the MOA results.
Note that the recent downward revision of the optical depth and event rate per star measurements from this survey~\citep{Sumi2016} does not affect our event rate per area results.

\subsection{Simulating {\it K2} Campaign 9 Observations of a Population of Free-Floating Planets}\label{params}

\begin{figure}
\includegraphics[width=\columnwidth]{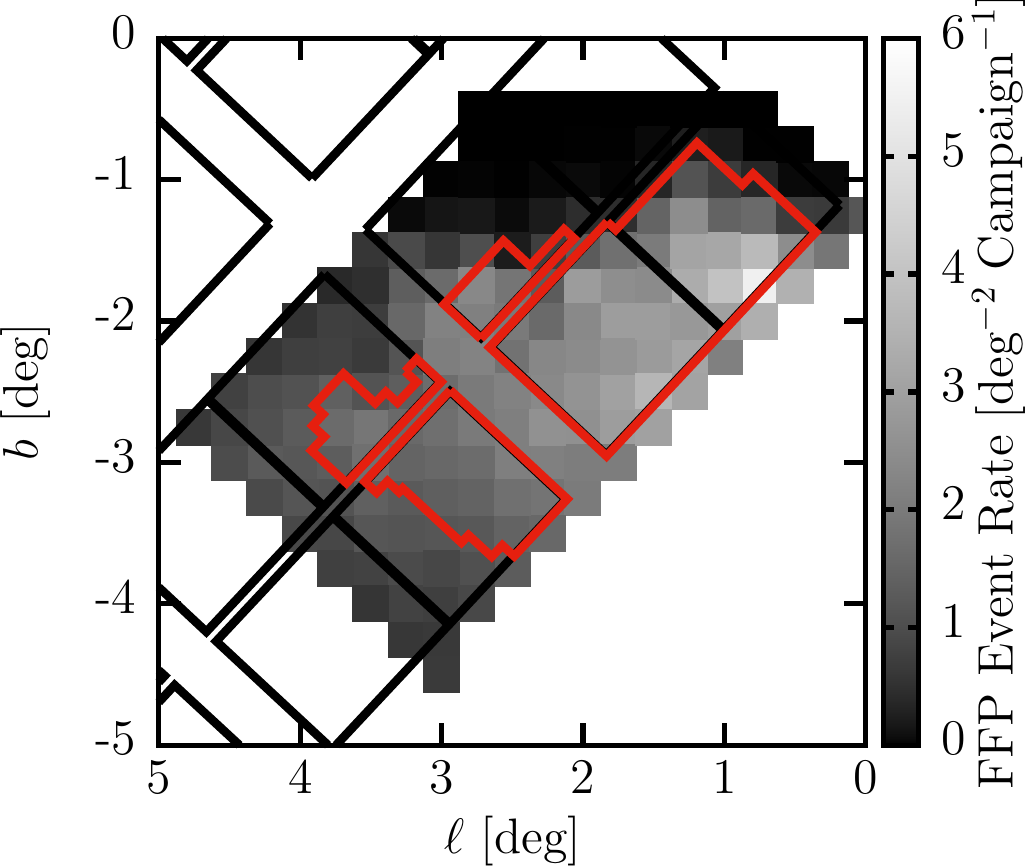}
\caption{Map of the expected free-floating planet event detection rate expected for \kt\ Campaign 9.
Black lines show the \ktcn\ footprint and red lines show the selected superstamp that will be downloaded.}
\label{map}
\end{figure}

\begin{figure}
\includegraphics[width=\columnwidth]{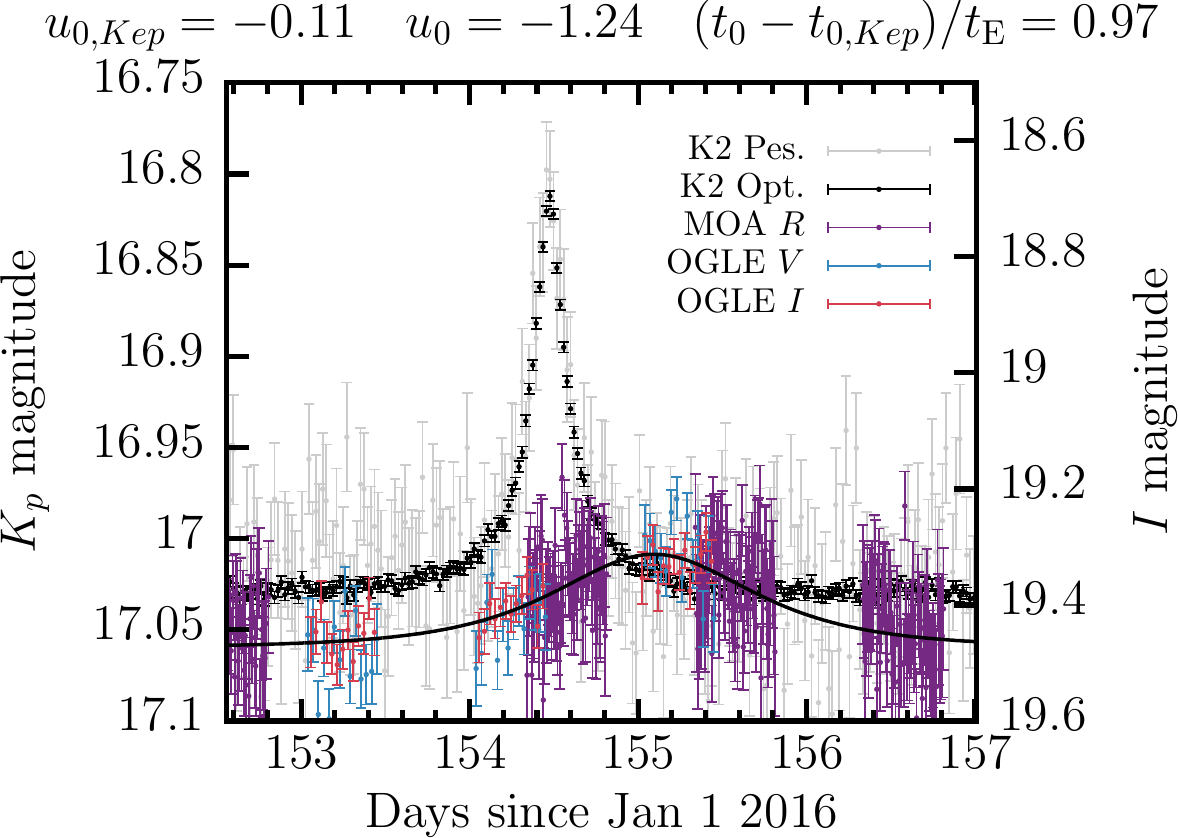}
\caption{Example of a simulated lightcurve.
Grey and black points, plotted against the left axis show pessimistic and optimistic \kt\ photometry, respectively.
Purple, blue and red points plotted against the right axis show MOA and OGLE $V$ and $I$ photometry, respectively.
Note that the $y$-scales and the baseline magnitudes are very different for the \kt\ and ground-based photometry, and that the event as seen from \kt\ is highly magnified but severely blended.}
\label{lc}
\end{figure}

We simulate the combination of the \kt\ Campaign 9 survey \citep[using the latest and final definition of the $3.74$~deg$^2$ superstamp chosen using the method of \citet{Poleski2015-rates} and shown in Figure~\ref{map}]{Henderson2015} with the ongoing high-cadence microlensing surveys of OGLE~\citep{Udalski2015-ogleiv} and MOA~\citep{Sako2007}.
At the time when we ran the simulations, the status of the KMTNet survey~\citep{KMTref, Henderson2014-kmt} for the 2016 season was uncertain, so we did not simulate it.
We therefore expect the ground-based sensitivity of our simulation to be very conservative.
Figure~\ref{lc} shows an example lightcurve.
The simulation parameters of the surveys and their detectors are listed in Table~\ref{simparams}.
Most of the parameters for \kt\ were taken from the \kepler\ {\it Instrument Handbook}~\citep{vanCleve2009}.
As the photometric precision it is possible to achieve with crowded field \kepler\ data is uncertain, we have simulated an optimisitic and pessimistic scenario for a systematic noise floor in the \kt\ data.
For the optimistic case, we simply add a $0.001$~fractional error in quadrature with the noise that is computed through our usual calculations that simulate CCD photometry~\citep{Penny2013}.
For the pessimistic case, we set an absolute noise floor of $45$~e$^{-}$~s$^{-1}$ independent of magnitude, which has been demonstrated in the crowded field of NGC 2158 in \kt\ Campaign 0 data~(Penny \& Stanek, 2016, in prep.).
In both cases we assume that the associated systematic noise terms account for any imperfect detrending.
See Section~\ref{discuss} for an assessment of the relative likelihood of each scenario.

Parameters for OGLE were taken from \citet{Udalski2015-ogleiv} or the OGLE website.\footnote{http://ogle.astrouw.edu.pl} 
The OGLE cadence was set by assuming OGLE would spend 50~percent of its time observing the superstamp, alternating between $V$ and $I$ filters.
The ${\sim}20$~minute combined cadence of OGLE $V$ and $I$ is similar to the actual OGLE cadence that will be used during the campaign, but the ratio of $V$ to $I$ exposures will be significantly reduced~(R. Poleski, private communication).
Parameters for MOA were taken from \citet{Sako2007}.
We assumed that MOA-$R$ magnitudes (a wide bandpass covering $R$ and $I$) were equal to $I$-band magnitudes.
Values for sky brightness, extinction coefficients and weather probabilities were taken from various appropriate observatory webpages.

We assumed that the population of FFPs is the same as that measured by \citet{Sumi2011}, namely that there are $1.9$ free-floating $1$-Jupiter-mass planets per main sequence star.
To implement this in our simulation, we replaced each of the lens stars in our simulation (including white dwarfs) with a FFP (adjusting the event rate weighting proportional to $\sqrt{M}$).
We then multiplied the number of detections by $1.9/(1+0.18)$, where the denominator is the sum of main sequence and white dwarf stars in the \citet{Sumi2011} model.
We drew the impact parameter $\uzkep$ from a uniform distribution in the range $-u_{\rm max}\le\uzkep<u_{\rm max}$, where $u_{\rm max}={\rm max}(1,2\rho)$ in order to include detectable events where the source is comparable or larger than the angular Einstein ring.
This input range does not include events where $\uzero<1$ (the impact parameter for Earth-bound observers) but $\uzkep>1$, which would potentially increase the number of characterizable events.

\subsection{Detection Criteria}

To determine if an event is detectable in \kt\ data we require an event cause a $\Delta\chi^2>200$ deviation from a flat lightcurve, and that $3$ or more consecutive data points deviate from the flat lightcurve by at least $3$-$\sigma$.
We consider an event to be detectable in ground-based data if it causes a $\Delta\chi^2>200$ deviation from a flat lightcurve.
For the Fisher matrix parameter uncertainty estimates, we consider a parameter to be ``measured'' if its fractional uncertainty is less than $1/3$ (i.e., $3$-$\sigma$).
Note that for parallax, for which there is a two-fold degeneracy in the magnitude of $\vecpie$~\citep{Refsdal1966,Gould1994}, the Fisher matrix estimate only characterizes one of the solutions. We discuss the impact of this degeneracy in Section~\ref{fourfold}.

\section{Results}\label{results}

\begin{table}
\caption{Predicted number of detections.}
\begin{tabular}{lcc}
\hline
\hline
 & Optimistic & Pessimistic \\

\hline
\kt\ detections  & 7.9 & 1.4\\
\kt\ and OGLE/MOA detections  & 5.1 & 1.2\\
$>3$-$\sigma$ $\pie$ measurements  & 3.9 & 1\\
$>3$-$\sigma$ $\rho$ measurements  & 1.1 & 0.44\\
$>3$-$\sigma$ mass measurements  & 0.98 & 0.42\\
\hline
\end{tabular}

\label{flowdown}
\end{table}

In contrast to the lengthy description of the simulations, the results can be presented much more concisely, and are summarized in Table~\ref{flowdown}.
For the optimistic \kt\ photometry case, if the FFP population is as described by \citet{Sumi2011} (1.9 Jupiter-mass planets per main sequence star), we expect that $\nhigh$ FFPs will be detected during Campaign 9.
Roughly $\fchihigh$~pecent of these have $\Delta\chi^2$ values between $200$ and $500$, with the rest having higher-significance detections.
For the pessimistic systematic noise scenario we expect $\nlow$ \kt\ detections, with $\fchilow$~percent between $\Delta\chi^2=200$ and $500$.
The probability of no \kt\ detections in the campaign is $\probnohigh$ for the optimistic simulation and $\probnolow$ for the pessimistic simulation.
The mean ratio of $\Delta\chi^2$ between the optimistic and pessimistic simulations is $\chiratio$, implying that there is significant room for improvement in photometry relative to that achieved by Penny \& Stanek (2016, in prep.).
The median $V$ and $I$ source magnitudes in the optimistic and pessimistic scenarios are $(V,I)=(\medVhigh,\medIhigh)$ and $(V,I)=(\medVlow,\medIlow)$, respectively.
These show that even with its large pixels, the \kepler\ spacecraft is extremely sensitive to microlensing events.

In the optimistic scenario, we predict that $\noglehigh$ of \kt's events will be detected in ground-based data as well, and in $\npiehigh$ of these it should be possible to make a $3$-$\sigma$ or better parallax measurement, up to the intrinsic four-fold degeneracy, which we discuss in Section~\ref{fourfold}.
Measurements of the angular Einstein radius will be more rare, with only $1.1$ predicted to have measurements of $\rho$.
This is not particularly surprising, since the typical $\theta_{\ast}$ is roughly an order of magnitude smaller than the typical $\thetae$ for Jovian mass lenses.
About $30$~percent of the $\rho$ measurements arise from events with sources of less than 1 solar radius. 
The mass and relative parallax of the lens can be measured by combining the microlensing parallax and angular Einstein radius as
\begin{equation}
M = \frac{\thetae}{\kappa\pie}; \quad \pirel = \pie\thetae.
\end{equation}
If we assume that a combination of a $3$-$\sigma$ $\pie$ measurement and a $3$-$\sigma$ $\rho$ measurement is sufficient to completely break the microlensing degeneracy and yield a mass and distance measurement, in nearly all cases where $\rho$ is measured, we find that parallax will also be measured.
So, our final expectation for the number of FFP mass measurements in the optimistic scenario is $\nmasshigh$, or a ${\sim} 60$ percent chance of a mass measurement assuming Poisson statistics.

Thankfully the attrition in the pessimistic scenario is a little less severe than in the optimistic scenario, principally because of the brighter sources to which \kt\ would be sensitive.
Out of $\nlow$~\kt\ detections, we expect $\npielow$ to yield a parallax and $\nrholow$ to yield a finite-source measurement.
We expect that $\nmasslow$ would yield a full mass measurement, or that there is a $34$~percent chance of a full mass measurement in the pessimistic scenario.

\section{Discussion}\label{discuss}

\subsection{Optimism versus Pessimism}

When examining the flow-down of detections to mass measurements in Table~\ref{flowdown}, one could be forgiven for being gloomy.
Before dispelling such gloom, however, we should address the relative levels of optimism and pessimism in our simulations.
The pessimistic noise floor we assume is equivalent to a 0.067~mag uncertainty at $K_p=18$, which one should recall is for a 30-minute integration on a 1-m class space telescope.
The photon noise contribution for such an integration at this magnitude, when also accounting for the expected ${\sim}17$ magnitudes per aperture of blended light, is $1.7$~mmag.
So, the pessimistic noise floor is a factor of ${\sim} 40$ higher than the photon noise at $K_p=18$.
This is not too surprising considering that Penny \& Stanek (2016, in prep.) first convolved the \kt\ data to give it a ${\sim}8$\arcsec\ PSF before performing difference imaging, and then measured the noise floor as the lower envelope of the lightcurve RMS when many lightcurves in the cluster core displayed blended variability.
Even if blending of variability is an issue in the Campaign 9 field, it will be possible to build an empirical model of it from ground-based data or, if the variable is periodic, remove it by folding~\citep[see, e.g.,][]{Wyrzykowski2006}.
Additionally, various techniques for extracting crowded field photometry are being explored by the \ktcn\ microlensing science team and others \citep[e.g.][who have achieved a noise floor of ${\sim}2$~mmag in crowded regions of NGC 2158]{Libralato2015}.
We therefore expect that the final \ktcn\ photometry will be much closer to our optimistic limit than our pessimistic limit.

\subsection{Characterization of the Free-Floating Planet Population}

So, given that we can expect multiple detections of short-timescale microlensing events, what can we learn from them? 
\citet{Henderson2016-char} investigated theoretically a similar flow-down of detections to characterization of FFPs in \ktcn\ as we have here, with an additional ingredient of a search for a possible stellar host to the lens. 
They concluded that it was unlikely to be possible to measure the combination of parallax, angular Einstein ring radius and rule out all possible stellar hosts down to the bottom of the main sequence in any single event. In other words, in all likelihood, it will not be possible to conclusively prove that any single candidate FFP event is both planetary and free-floating. Our results agree with this assessment, but we feel that \citet{Henderson2016-char} missed an important opportunity to consider what one can learn from partial information for many FFP candidate events. Without doing so, one could be forgiven for drawing a pessimistic conclusion from the discussion in \citet{Henderson2016-char}. The conclusions we draw from our simulations is much more optimistic.

We have shown that, with reasonable assumptions about the achievable \kt\ precision, FFP events will be detectable in the \kt\ data.
Should the short timescale events all be caused by FFPs, it will be possible to measure parallaxes in over half of the detected events, and for many more if they are caused by stellar-mass lenses.
In the remainder of FFP events, the microlensing event will not be seen from the ground, sometimes due to poor weather, but other times because the large parallax has caused the impact parameter from the ground to be too large for an event to be detected.
For the latter scenario then, it will be possible to place a lower limit on the parallax, which with a reasonable assumption of the kinematics and density distribution of the Galaxy can be translated into a statistical upper limit on the mass.

\begin{figure}
\includegraphics[width=\columnwidth]{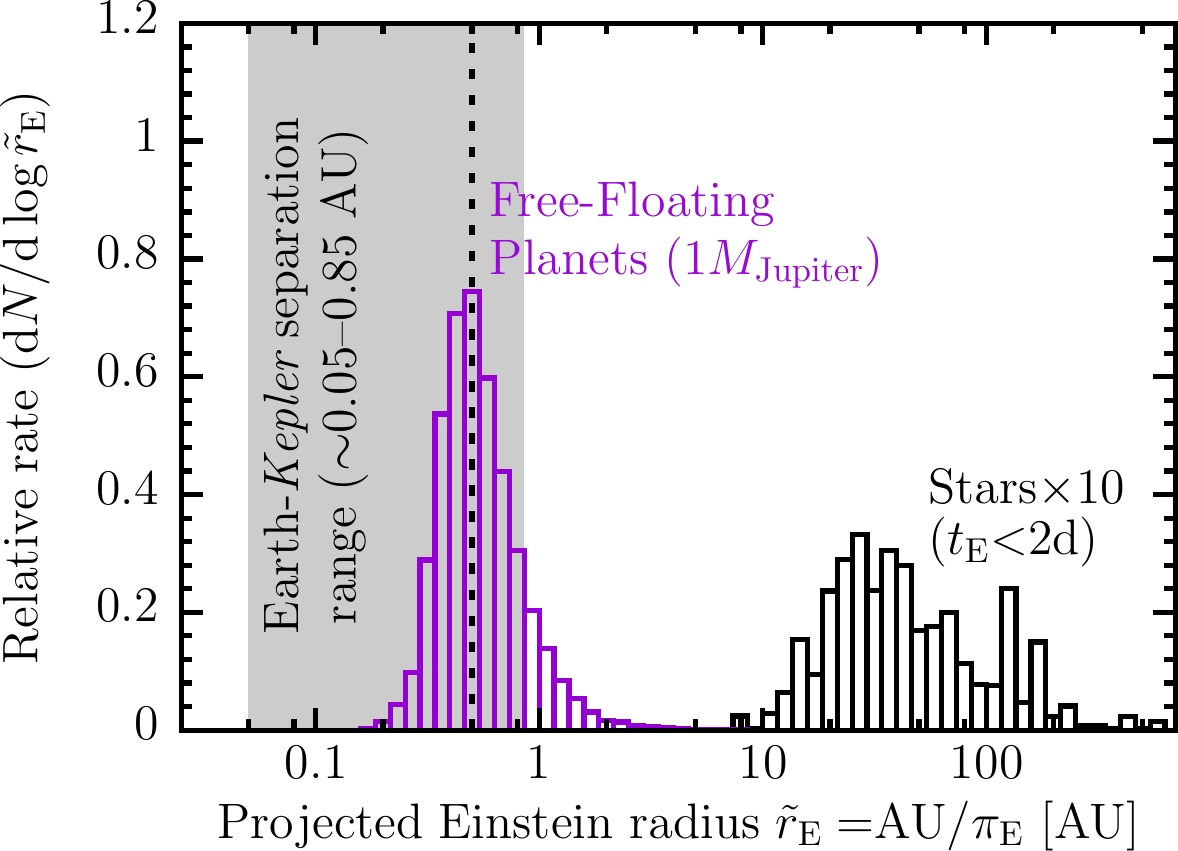}
\caption{Distribution of projected Einstein radii for $1$-$\mjup$ FFPs (purple) and for stars that cause microlensing events shorter than $2$~days (black) detected in our simulations; note that the normalization for the stars distribution has been multiplied by 10 to make details visible.
These distributions are compared to the range projected sepraration between Earth and \kepler\ (shaded region).
The dashed line shows the projected separation at the midpoint of \ktcn.
Short-timescale events caused by stars will always have small parallaxes ${\sim}0.1$--$0.01$, and so any event detectable from \kt\ should be detectable from the ground.
The corollary is that any event that is only detectable from one location is almost certainly caused by a planetary-mass lens.}
\label{reproj}
\end{figure}

In Figure~\ref{reproj}, we show the distribtion of projected Einstein radii, $\reproj=$AU$/\pie$, for FFPs relative to the range of projected Earth-\kt\ baselines $\dbase$ that are possible during the campaign.
It can be seen that in many events $\reproj$ will be smaller than the baseline, meaning that it is probable for those events (though not certain) that the event will only be detectable from one location.
However, we also plot the distribution of $\reproj$ for stellar-mass lenses with timescales less than $2$~days (i.e., FFP impostors).
For these events, $\reproj$ is \emph{always} larger than the baseline, so it should always be possible to detect an event that \kepler\ sees from the ground as well, provided that the weather is good.
Therefore, for \emph{all} the events with suitable ground-based coverage that are detectable from \kt, it will be possible to statistically infer with a high degree of confidence whether the mass of the lensing object is roughly planetary or roughly stellar.
Angular Einstein ring radius measurements are not required for this inference.
Note also that the normalization of the stellar distribution in Figure~\ref{reproj} was multiplied by $10$ in order for the distribution to be visible, so that even without a parallax measurement or lower limit, a short timescale, $\tein<2$~days already strongly implies a planetary mass.

\begin{figure}
\includegraphics[width=\columnwidth]{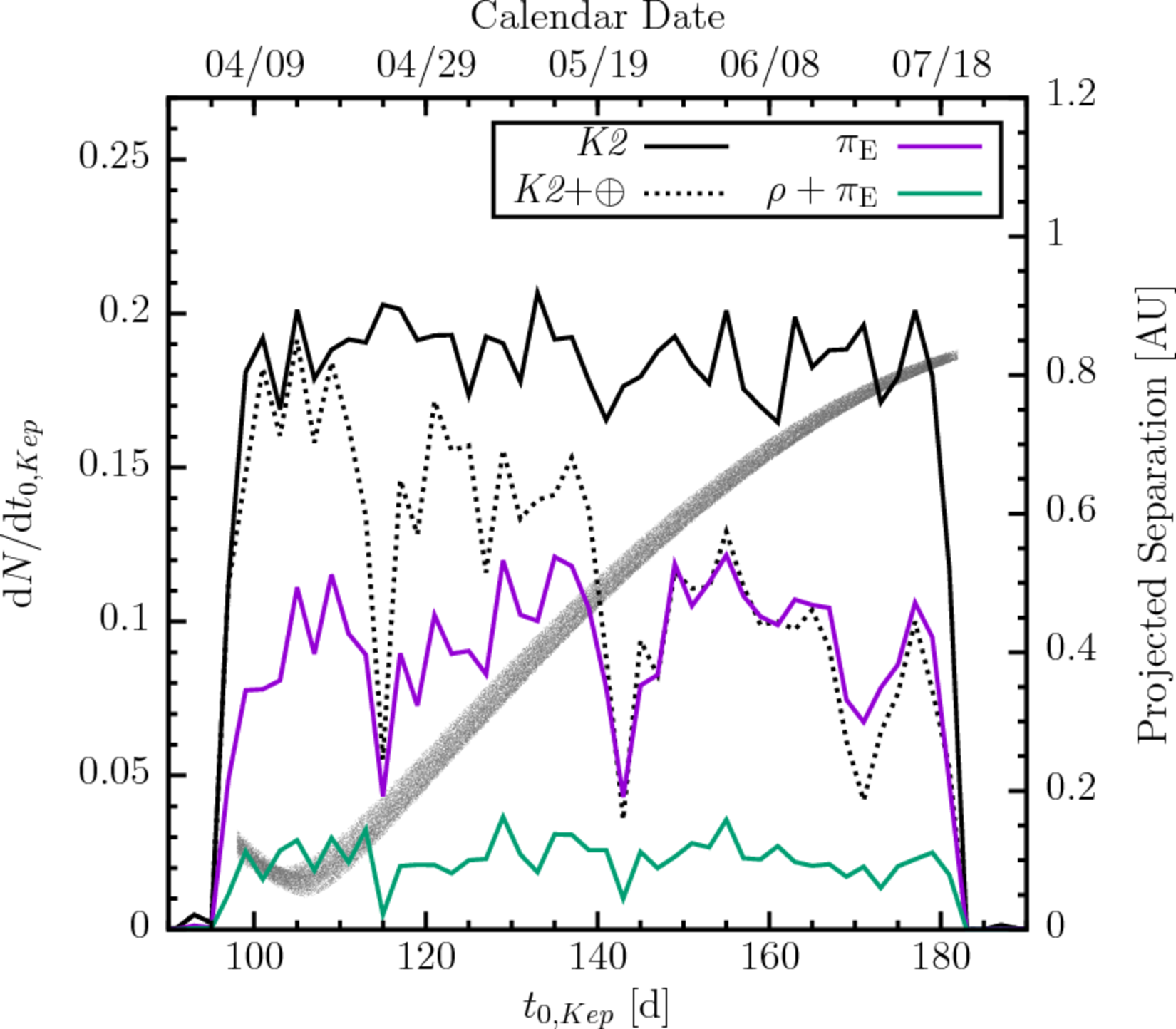}
\caption{The distribution of event peak times $\tzkep$ for events detectable in only \kt\ data (solid black line) and in both \kt\ and ground-based data (dotted black line). 
Purple and green lines show the distribution of $3$-$\sigma$ parallax measurements and parallax plus finite source measurements, respectively.
The gray swath shaped like a check mark and plotted against the right axis, is the distribution of projected separations between \kt\ and Earth for simulated events; the vertical spread is caused by events with a range of ecliptic latitudes.
Even at the end of the campaign when the projected separation is larger than the typical FFP projected Einstein radius, a significant fraction of events seen by \kt\ can be detected from Earth, after accounting for the 3 dips in ground-based detections that occur when the moon is near the bulge fields and no observations are simulated.}
\label{t0dist}
\end{figure}

So, following \ktcn\ we will know to a good degree of accuracy the fraction of the short-timescale events that are caused by planetary-mass objects.
Can we also infer whether the population is truly free-floating or is just loosely bound to stars unseen through lensing? \citet{Henderson2016-char} considered success as the ability to rule out all stellar hosts down to the hydrogen burning limit by prompt adaptive optics (AO) imaging of events while the source and lens are still together, and argued that it was only possible for nearby events, within $2$~kpc.
These events would have the smallest $\reproj$ and so could only have a parallax measurement in the early part of \ktcn\ when the Earth-\kt\ baseline was also small.
However, if one instead is trying to answer the question of whether the \emph{population} is bound or not, one can use a series of partial exclusions to make inferences about the population.
In Figure~\ref{t0dist}, we show that it will be possible to detect FFPs in both \kt\ and ground-based data throughout the campaign, although with an efficiency that falls gradually throughout the campaign, down to ${\sim}50$~percent at the end of the season compared to nearly $100$~percent at the start. 
Interestingly, however, we find that the distribution of parallax measurements and full-mass measurements is essentially uniform throughout the campaign, likely due to a trade-off between the size of the parallax effect (which decreases with as the campaign goes on) and the nightly duration of bulge visibility (which increases as the campaign progresses).
The falling trend in ground-based detections of \kt\ events is due to the \kt-Earth projected baseline becoming gradually larger than the typical projected Einstein-ring radius.
The three dips in ground-based detections occur at the times when the moon is passing close to the bulge fields, and so there are no ground-based observations.

\citet{Henderson2016-char} argued that it should be possible, in events with main sequence or subgiant sources, to rule out FFP hosts down to $0.25\msun$ in the bulge.
Assuming a \citet{Kroupa2001} IMF, this rules out 29~percent of the IMF by number, if we assume that it runs from $0.08$ to $1.0\msun$, with more massive stars in the bulge having already evolved.
If \ktcn\ detects 8 FFP candidates and all of them are in fact bound planets, we can expect $0.29\times 8=2.35$ of them to have detectable hosts; there is therefore a $10$~percent probability that a survey of all 8 FFPs would yield no detected hosts.
However, once we have classified the population as largely planetary using parallax measurements and lower limits, we can bring to bear the full statistical power of \emph{all} short-timescale events to ask whether they are consistent with a bound population.
\citet{Sumi2011} found $10$ events with $\tein<2$~days in 2 years of MOA data from 2006-2007; with the subsequent upgrade of the OGLE survey~\citep{Udalski2015-ogleiv} and the advent of KMTNet~\citep{KMTref,Henderson2014-kmt}, the number of short timescale events suitable for such AO characterization observations should approach $100$ in the coming few years.
Note however, that in order to place the strongest constraints on potential hosts using prompt AO follow-up, it is necessary to measure the source flux in the infrared during the FFP event, which would require a complimentary near-IR microlensing survey with roughly daily cadence, or a rapid response TOO program.
Such a sample would be easily large enough to rule out a population of stellar hosts, even accounting for the incidence of binary companions to source stars.
Less-prompt high-resolution follow-up can place more stringent limits on the presence of a host once it has moved away from the source star of the event~\citep[e.g.][]{Bennett2007,Henderson2015-ao}.
Similar constraints on host stars could be placed over a wide range of FFP masses by pre-covering a significant fraction of the WFIRST~\citep{Spergel2015} microlensing field with HST imaging (e.g., by covering an area of the bulge in similar manner to the PHAT survey.

\subsection{Four-fold Degeneracy}\label{fourfold}

So far, we have not paid much attention to the four-fold degeneracy that affects satellite parallax measurements from just two locations.
In this section we consider how it affects our conclusions.
The degeneracy arises from the inability to tell upon which side of the lens the source passes when observed from each observatory~\citep[]{Refsdal1966, Gould1994}.
This means that $\Delta\uzero$ in equation~\ref{dtdu} can take one of two values, and the sign of each of these is also undetermined.
The sign only affects the direction of the parallax vector and not its magnitude, so is not relevent to the estimation of the lens mass, but the magnitude of $\Delta\uzero$ is.

In the case of a full solution of the lensing event, with both finite source effects and parallax measured, the magnitude of $\pie$ will be approximately the same for the two degenerate solutions, because either $\uzero$ or $\uzkep$ will be close to zero, and difference between the degenerate values of $\Delta\uzero$ will be of the same order of magnitude as the uncertainty on $\uzero$ or $\uzkep$, which will be of the order of the source size.

In the case where the event is only detected from \kepler, and assuming that there is sufficient coverage from the ground, it will only be possible to place a lower limit on $|\uzero|$, which will imply a lower limit on $|\Delta\uzero|$ equal to the lower limit on $|\uzero|$ minus $|\uzkep|$.
This assumes the case where the source passes on the same side of the lens for both \kepler\ and Earth; the degenerate case with the source passing on opposite sides, and which holds $50$~percent probability, has a larger lower limit on $\Delta\uzero$ and $\pie$.
For 75~percent of the events \kepler\ detects, $\uzkep<0.5$, and we found in the simulations that ground-based observatories maintain detection efficiency to beyond $|\uzero|>1$, so for most such events it will be possible rule out $\uzero\sim\uzkep$.
As stellar-mass lenses will always have small values of $\Delta\uzero$ (the 99th percentile of $\Delta\uzero$ for stellar-mass lenses with $\tein<2$~days is $0.17$), this means that for virtually all events that \kepler\ detects but are not seen from Earth, it will be possible to conclusively rule out stellar-mass lenses, regardless of the parallax degeneracy.

Finally, we are left to consider the cases without finite source effects and with a detection of the event from both \kepler\ and Earth.
For events genuinely caused by FFPs, many will have values of $\Delta\tzero/\tekep$ large enough to render the degeneracy moot in the separation of stellar-mass from planetary-mass lenses.
If $\Delta\tzero/\tekep\lesssim 0.2$ though, we must consider the degeneracy.
For genuine FFP events where $\Delta\tzero/\tekep$, is small compared to $\pie D_{\perp}$, then $\Delta\uzero\simeq \pie D_{\perp}$, which will be guaranteed to cause a large difference in $\uzero$ if the lens passes on the same side of the lens, and in almost all cases when the source passes on the opposite side of the lens.
The chance that both $\Delta\tzero/\tekep\sim 0$ and $\Delta\uzero\simeq 2|\uzkep|$ is very small if the lens is a FFP.
However, if the lens is a star, $\pie D_{\perp}$ will be small and hence $\Delta\uzero$ will always be small, so we would conclude that probably such an event would have a stellar-mass lens.
This is essentially a restatement of the ``Rich argument,'' that fine tuning is required in order for an event with large parallax to have both small $\Delta\tzero/\tekep$ and $\Delta\uzero$~\citep[see][for a full description and quantification of the probabilities]{CalchiNovati2015}, but in the context of constraining the lens' mass regime rather than its Galactic location.

\subsection{Impact of {\it Kepler}'s Emergency Mode}

The simulations we present were performed assuming a continuous 84 day \kt\ campaign with no mid-campaign break. The loss of the beginning of \ktcn\ due to a spacecraft emergency mode will obviously reduce the number of FFP events we can expect to detect. \kepler\ began observing again on 22nd~April,\footnote{https://www.nasa.gov/feature/ames/kepler/mission-manager-update-kepler-recovered-and-returned-to-the-k2-mission} which reduces the data-collecting duration of \ktcn\ to $67$ days, or $80$~percent of the duration we simulated. As can be seen from Figure~\ref{t0dist}, \kepler\ detects events uniformly in time, and the rate of characterizable events behaves in the same way, so all the predictions in Table~\ref{flowdown} can be reduced by $20$~percent, except the second line, which would be reduced by a slightly larger amount.

\section{Conclusions}\label{conclusions}

We have performed detailed simulations of \kt\ Campaign 9 and the accompanying ground-based observations to predict the yield of FFP detections if a population of $1$-$\mjup$ FFPs as infered by \citet{Sumi2011} exists.
We expect that the nominal \kt\ campaign would have detected ${\sim} 7.9$ FFP events (under our optimistic, but more likely realistic assumptions), with most of them also being detected in ground-based observations, enabling parallax measurements.
Even for the events not detected from the ground it should be possible to place upper limits on the mass of the FFP candidate given sufficient ground-based coverage and a reasonable assumption of Galactic model.
We argue that prompt AO observations of a large sample of short-timescale events, combined with the parallax results of \ktcn\ should be able to conclusively show that the population of short-timescale events discovered by \citet{Sumi2011} is a population of genuinely free-floating, genuinely planetary-mass objects, or prove that this is not the case.

\acknowledgments
MTP would like to thank Calen Henderson and Yossi Shvartzvald for the discussions that provided the impetus to write, and for detailed comments on the paper.
This work was performed under contract with the California Institute of Technology (Caltech)/Jet Propulsion Laboratory (JPL) funded by NASA through the Sagan Fellowship Program executed by the NASA Exoplanet Science Institute.
Work by MTP and BSG was supported by NASA grant NNX16AC62G.
NJR is a Royal Society of New Zealand Rutherford Discovery Fellow.

\software{GPC: General Polygon Clipper library ascl:1512.006, MATLAB package for astronomy and astrophysics ascl:1407.005}

\bibliographystyle{aasjournal}
\bibliography{libraryshort}

\begin{thebibliography}{}
\expandafter\ifx\csname natexlab\endcsname\relax\def\natexlab#1{#1}\fi

\bibitem[{Awiphan {et~al.}(2016)Awiphan, Kerins, \& Robin}]{Awiphan2016}
Awiphan, S., Kerins, E., \& Robin, A.~C. 2016, MNRAS, 456, 1666

\bibitem[{Bennett {et~al.}(2007)Bennett, Anderson, \& Gaudi}]{Bennett2007}
Bennett, D.~P., Anderson, J., \& Gaudi, B.~S. 2007, ApJ, 660, 781

\bibitem[{Bowler {et~al.}(2015)Bowler, Liu, Shkolnik, \& Tamura}]{Bowler2015}
Bowler, B.~P., Liu, M.~C., Shkolnik, E.~L., \& Tamura, M. 2015, ApJS, 216, 7

\bibitem[{Bozza {et~al.}(2016)Bozza, Shvartzvald, Udalski, Novati, Bond, Han,
  Hundertmark, Poleski, Pawlak, Szyma{\'{n}}ski, Skowron, Mr{\'{o}}z,
  Koz{\l}owski, Wyrzykowski, Pietrukowicz, Soszy{\'{n}}ski, Ulaczyk, Beichman,
  Bryden, Carey, Fausnaugh, Gaudi, Gould, Henderson, Pogge, Wibking, Yee, Zhu,
  Abe, Asakura, Barry, Bennett, Bhattacharya, Donachie, Freeman, Fukui, Hirao,
  Inayama, Itow, Koshimoto, Li, Ling, Masuda, Matsubara, Muraki, Nagakane,
  Nishioka, Ohnishi, Oyokawa, Rattenbury, Saito, Sharan, Sullivan, Sumi,
  Suzuki, Tristram, Wakiyama, Yonehara, Choi, Park, Jung, Shin, Albrow, Park,
  Kim, Lee, Cha, Kim, Lee, Dominik, J{\o}rgensen, Andersen, Bramich, Burgdorf,
  Ciceri, D'Ago, Evans, Jaimes, Gu, Hinse, Kains, Kerins, Korhonen, Kuffmeier,
  Mancini, Popovas, Rabus, Rahvar, Rasmussen, Scarpetta, Skottfelt, Snodgrass,
  Southworth, Surdej, Unda-Sanzana, von Essen, Wang, Wertz, Maoz, Friedmann, \&
  Kaspi}]{Bozza2016}
Bozza, V., Shvartzvald, Y., Udalski, A., {et~al.} 2016, ApJ, 820, 79

\bibitem[{Brown {et~al.}(2011)Brown, Latham, Everett, \& Esquerdo}]{Brown2011}
Brown, T.~M., Latham, D.~W., Everett, M.~E., \& Esquerdo, G.~A. 2011, AJ, 142,
  112

\bibitem[{Bryson {et~al.}(2010)Bryson, Tenenbaum, Jenkins, Chandrasekaran,
  Klaus, Caldwell, Gilliland, Haas, Dotson, Koch, \& Borucki}]{Bryson2010}
Bryson, S.~T., Tenenbaum, P., Jenkins, J.~M., {et~al.} 2010, ApJ, 713, L97

\bibitem[{Calamida {et~al.}(2015)Calamida, Sahu, Casertano, Anderson, Cassisi,
  Gennaro, Cignoni, Brown, Kains, Ferguson, Livio, Bond, Buonanno, Clarkson,
  Ferraro, Pietrinferni, Salaris, \& Valenti}]{Calamida2015}
Calamida, A., Sahu, K.~C., Casertano, S., {et~al.} 2015, ApJ, 810, 8

\bibitem[{{Calchi Novati} {et~al.}(2015){Calchi Novati}, Gould, Udalski,
  Menzies, Bond, Shvartzvald, Street, Hundertmark, Beichman, Yee, Carey,
  Poleski, Skowron, Koz{\l}owski, Mr{\'{o}}z, Pietrukowicz, Pietrzy{\'{n}}ski,
  Szyma{\'{n}}ski, Soszy{\'{n}}ski, Ulaczyk, Wyrzykowski, Albrow, Beaulieu,
  Caldwell, Cassan, Coutures, Danielski, {Dominis Prester}, Donatowicz,
  Lon{\v{c}}ari{\'{c}}, McDougall, Morales, Ranc, Zhu, Abe, Barry, Bennett,
  Bhattacharya, Fukunaga, Inayama, Koshimoto, Namba, Sumi, Suzuki, Tristram,
  Wakiyama, Yonehara, Maoz, Kaspi, Friedmann, Bachelet, {Figuera Jaimes},
  Bramich, Tsapras, Horne, Snodgrass, Wambsganss, Steele, Kains, Bozza,
  Dominik, J{\o}rgensen, Alsubai, Ciceri, D'Ago, Haugb{\o}lle, Hessman, Hinse,
  Juncher, Korhonen, Mancini, Popovas, Rabus, Rahvar, Scarpetta, Schmidt,
  Skottfelt, Southworth, Starkey, Surdej, Wertz, Zarucki, Gaudi, Pogge, \&
  DePoy}]{CalchiNovati2015}
{Calchi Novati}, S., Gould, A., Udalski, A., {et~al.} 2015, ApJ, 804, 20

\bibitem[{Cardelli {et~al.}(1989)Cardelli, Clayton, \& Mathis}]{Cardelli1989}
Cardelli, J.~A., Clayton, G.~C., \& Mathis, J.~S. 1989, ApJ, 345, 245

\bibitem[{Chabrier(2003)}]{Chabrier2003}
Chabrier, G. 2003, PASP, 115, 763

\bibitem[{Clanton \& Gaudi(2014)}]{Clanton2014a}
Clanton, C., \& Gaudi, B.~S. 2014, ApJ, 791, 91

\bibitem[{Clanton \& Gaudi(2016)}]{Clanton2016}
---. 2016, ApJ, 819, 125

\bibitem[{Dong {et~al.}(2007)Dong, Udalski, Gould, Reach, Christie, Boden,
  Bennett, Fazio, Griest, Szyma{\'{n}}ski, Kubiak, Soszy{\'{n}}ski,
  Pietrzy{\'{n}}ski, Szewczyk, Wyrzykowski, Ulaczyk, Wieckowski,
  Paczy{\'{n}}ski, DePoy, Pogge, Preston, Thompson, \& Patten}]{Dong2007}
Dong, S., Udalski, A., Gould, A., {et~al.} 2007, ApJ, 664, 862

\bibitem[{Durkan {et~al.}(2016)Durkan, Janson, \& Carson}]{Durkan2016}
Durkan, S., Janson, M., \& Carson, J. 2016, eprint arXiv:1604.00859

\bibitem[{Gould(1994)}]{Gould1994}
Gould, A. 1994, ApJ, 421, L75

\bibitem[{Gould(2000)}]{Gould2000}
---. 2000, ApJ, 542, 785

\bibitem[{Gould(2003)}]{Gould2003}
---. 2003, eprint arXiv:astro-ph/0310577

\bibitem[{Gould \& Horne(2013)}]{Gould2013-whitepaper}
Gould, A., \& Horne, K. 2013, eprint arXiv:1306.2308

\bibitem[{Gwyn(2008)}]{Gwyn2008}
Gwyn, S. D.~J. 2008, PASP, 120, 212

\bibitem[{Henderson(2015)}]{Henderson2015-ao}
Henderson, C.~B. 2015, ApJ, 800, 58

\bibitem[{Henderson {et~al.}(2014)Henderson, Gaudi, Han, Skowron, Penny, Nataf,
  \& Gould}]{Henderson2014-kmt}
Henderson, C.~B., Gaudi, B.~S., Han, C., {et~al.} 2014, ApJ, 794, 52

\bibitem[{Henderson \& Shvartzvald(2016)}]{Henderson2016-char}
Henderson, C.~B., \& Shvartzvald, Y. 2016, eprint arXiv:1603.05249

\bibitem[{Henderson {et~al.}(2015)Henderson, Poleski, Penny, Street, Bennett,
  Hogg, Gaudi, Zhu, Barclay, Barentsen, Howell, Mullally, Udalski,
  Szyma{\'{n}}ski, Skowron, Mr{\'{o}}z, Koz{\l}owski, Wyrzykowski,
  Pietrukowicz, Soszy{\'{n}}ski, Ulaczyk, Pawlak, Sumi, Abe, Asakura, Barry,
  Bhattacharya, Bond, Donachie, Freeman, Fukui, Hirao, Itow, Koshimoto, Li,
  Ling, Masuda, Matsubara, Muraki, Nagakane, Ohnishi, Oyokawa, Rattenbury,
  Saito, Sharan, Sullivan, Tristram, Yonehara, Bachelet, Bramich, Cassan,
  Dominik, {Figuera Jaimes}, Horne, Hundertmark, Mao, Ranc, Schmidt, Snodgrass,
  Steele, Tsapras, Wambsganss, Bozza, Burgdorf, J{\o}rgensen, {Calchi Novati},
  Ciceri, D'Ago, Evans, Hessman, Hinse, Husser, Mancini, Popovas, Rabus,
  Rahvar, Scarpetta, Skottfelt, Southworth, Unda-Sanzana, Bryson, Caldwell,
  Haas, Larson, McCalmont, Packard, Peterson, Putnam, Reedy, Ross, {Van Cleve},
  Akeson, Batista, Beaulieu, Beichman, Bryden, Ciardi, Cole, Coutures,
  Foreman-Mackey, Fouqu{\'{e}}, Friedmann, Gelino, Kaspi, Kerins, Korhonen,
  Lang, Lee, Lineweaver, Maoz, Marquette, Mogavero, Morales, Nataf, Pogge,
  Santerne, Shvartzvald, Suzuki, Tamura, Tisserand, \& Wang}]{Henderson2015}
Henderson, C.~B., Poleski, R., Penny, M., {et~al.} 2015, eprint
  arXiv:1512.09142

\bibitem[{Howell {et~al.}(2014)Howell, Sobeck, Haas, Still, Barclay, Mullally,
  Troeltzsch, Aigrain, Bryson, Caldwell, Chaplin, Cochran, Huber, Marcy,
  Miglio, Najita, Smith, Twicken, \& Fortney}]{Howell2014}
Howell, S.~B., Sobeck, C., Haas, M., {et~al.} 2014, PASP, 126, 398

\bibitem[{Johnson \& Soderblom(1987)}]{Johnson1987}
Johnson, D. R.~H., \& Soderblom, D.~R. 1987, AJ, 93, 864

\bibitem[{Kerins {et~al.}(2009)Kerins, Robin, \& Marshall}]{Kerins2009}
Kerins, E., Robin, A.~C., \& Marshall, D.~J. 2009, MNRAS, 396, 1202

\bibitem[{Kim {et~al.}(2010)Kim, Park, Lee, Yuk, Han, O'Brien, Gould, Lee, \&
  Kim}]{KMTref}
Kim, S.-L., Park, B.-G., Lee, C.-U., {et~al.} 2010, Proceedings of the SPIE,
  7733, 3F

\bibitem[{Krisciunas \& Schaefer(1991)}]{Krisciunas1991}
Krisciunas, K., \& Schaefer, B.~E. 1991, PASP, 103, 1033

\bibitem[{Kroupa(2001)}]{Kroupa2001}
Kroupa, P. 2001, MNRAS, 322, 231

\bibitem[{Leinert {et~al.}(1998)Leinert, Bowyer, Haikala, Hanner, Hauser,
  Levasseur-Regourd, Mann, Mattila, Reach, Schlosser, Staude, Toller, Weiland,
  Weinberg, \& Witt}]{Leinert1998}
Leinert, C., Bowyer, S., Haikala, L.~K., {et~al.} 1998, Astronomy and
  Astrophysics Supplement Series, 127, 1

\bibitem[{Libralato {et~al.}(2015)Libralato, Bedin, Nardiello, \&
  Piotto}]{Libralato2015}
Libralato, M., Bedin, L.~R., Nardiello, D., \& Piotto, G. 2015, MNRAS, 456,
  1137

\bibitem[{Lodieu(2013)}]{Lodieu2013}
Lodieu, N. 2013, MNRAS, 431, 3222

\bibitem[{Marshall {et~al.}(2006)Marshall, Robin, Reyl{\'{e}}, Schultheis, \&
  Picaud}]{Marshall2006}
Marshall, D.~J., Robin, A.~C., Reyl{\'{e}}, C., Schultheis, M., \& Picaud, S.
  2006, A\&A, 453, 635

\bibitem[{Moffat(1969)}]{Moffat1969}
Moffat, A. F.~J. 1969, A\&A, 3, 455

\bibitem[{Nataf {et~al.}(2013)Nataf, Gould, Fouqu{\'{e}}, Gonzalez, Johnson,
  Skowron, Udalski, Szyma{\'{n}}ski, Kubiak, Pietrzy{\'{n}}ski,
  Soszy{\'{n}}ski, Ulaczyk, Wyrzykowski, \& Poleski}]{Nataf2013}
Nataf, D.~M., Gould, A., Fouqu{\'{e}}, P., {et~al.} 2013, ApJ, 769, 88

\bibitem[{Ofek(2014)}]{Ofek2014}
Ofek, E.~O. 2014, Astrophysics Source Code Library

\bibitem[{{Pe{\~{n}}a Ram{\'{i}}rez} {et~al.}(2012){Pe{\~{n}}a Ram{\'{i}}rez},
  B{\'{e}}jar, {Zapatero Osorio}, Petr-Gotzens, \&
  Mart{\'{i}}n}]{PenaRamirez2012}
{Pe{\~{n}}a Ram{\'{i}}rez}, K., B{\'{e}}jar, V. J.~S., {Zapatero Osorio},
  M.~R., Petr-Gotzens, M.~G., \& Mart{\'{i}}n, E.~L. 2012, ApJ, 754, 30

\bibitem[{Penny {et~al.}(2013)Penny, Kerins, Rattenbury, Beaulieu, Robin, Mao,
  Batista, {Calchi Novati}, Cassan, Fouque, McDonald, Marquette, Tisserand, \&
  {Zapatero Osorio}}]{Penny2013}
Penny, M.~T., Kerins, E., Rattenbury, N., {et~al.} 2013, MNRAS, 434, 2

\bibitem[{Pfyffer {et~al.}(2015)Pfyffer, Alibert, Benz, \&
  Swoboda}]{Pfyffer2015}
Pfyffer, S., Alibert, Y., Benz, W., \& Swoboda, D. 2015, Astronomy {\&}
  Astrophysics, 579, A37

\bibitem[{Poleski(2015)}]{Poleski2015-rates}
Poleski, R. 2015, MNRAS, 455, 3656

\bibitem[{Poleski {et~al.}(2015)Poleski, Zhu, Christie, Udalski, Gould,
  Bachelet, Skottfelt, {Calchi Novati}, Szyma{\'{n}}ski, Soszy{\'{n}}ski,
  Pietrzy{\'{n}}ski, Wyrzykowski, Ulaczyk, Pietrukowicz, Koz{\l}owski, Skowron,
  Mr{\'{o}}z, Pawlak, Beichman, Bryden, Carey, Fausnaugh, Gaudi, Henderson,
  Pogge, Shvartzvald, Wibking, Yee, Beatty, Eastman, Drummond, Friedmann,
  Henderson, Johnson, Kaspi, Maoz, McCormick, McCrady, Natusch, Ngan, Porritt,
  Relles, Sliski, Tan, Wittenmyer, Wright, Street, Tsapras, Bramich, Horne,
  Snodgrass, Steele, Menzies, {Figuera Jaimes}, Wambsganss, Schmidt, Cassan,
  Ranc, Mao, Bozza, Dominik, Hundertmark, J{\o}rgensen, Andersen, Burgdorf,
  Ciceri, D'Ago, Evans, Gu, Hinse, Kains, Kerins, Korhonen, Kuffmeier, Mancini,
  Popovas, Rabus, Rahvar, Rasmussen, Southworth, Surdej, Unda-Sanzana, Verma,
  von Essen, Wang, \& Wertz}]{Poleski2015-spitzer}
Poleski, R., Zhu, W., Christie, G.~W., {et~al.} 2015, eprint arXiv:1512.08520

\bibitem[{Quanz {et~al.}(2012)Quanz, Lafreni{\`{e}}re, Meyer, Reggiani, \&
  Buenzli}]{Quanz2012}
Quanz, S.~P., Lafreni{\`{e}}re, D., Meyer, M.~R., Reggiani, M.~M., \& Buenzli,
  E. 2012, Astronomy {\&} Astrophysics, 541, A133

\bibitem[{Raghavan {et~al.}(2010)Raghavan, McAlister, Henry, Latham, Marcy,
  Mason, Gies, White, \& ten Brummelaar}]{Raghavan2010}
Raghavan, D., McAlister, H.~A., Henry, T.~J., {et~al.} 2010, ApJS, 190, 1

\bibitem[{Refsdal(1966)}]{Refsdal1966}
Refsdal, S. 1966, MNRAS, 134, 315

\bibitem[{Reggiani {et~al.}(2016)Reggiani, Meyer, Chauvin, Vigan, Quanz,
  Biller, Bonavita, Desidera, Delorme, Hagelberg, Maire, Boccaletti, Beuzit,
  Buenzli, Carson, Covino, Feldt, Girard, Gratton, Henning, Kasper, Lagrange,
  Mesa, Messina, Montagnier, Mordasini, Mouillet, Schlieder, Segransan,
  Thalmann, \& Zurlo}]{Reggiani2016}
Reggiani, M., Meyer, M.~R., Chauvin, G., {et~al.} 2016, Astronomy {\&}
  Astrophysics, 586, A147

\bibitem[{Robin {et~al.}(2003)Robin, Reyl�, Derri�re, \&
  Picaud}]{Robin2003}
Robin, A.~C., Reyl�, C., Derri�re, S., \& Picaud, S. 2003, A\&A, 409, 523

\bibitem[{Sako {et~al.}(2007)Sako, Sekiguchi, Sasaki, Okajima, Abe, Bond,
  Hearnshaw, Itow, Kamiya, Kilmartin, Masuda, Matsubara, Muraki, Rattenbury,
  Sullivan, Sumi, Tristram, Yanagisawa, \& Yock}]{Sako2007}
Sako, T., Sekiguchi, T., Sasaki, M., {et~al.} 2007, ExA, 22, 51

\bibitem[{Salpeter(1955)}]{Salpeter1955}
Salpeter, E.~E. 1955, ApJ, 121, 161

\bibitem[{Shvartzvald {et~al.}(2015)Shvartzvald, Udalski, Gould, Han, Bozza,
  Friedmann, Hundertmark, Beichman, Bryden, Novati, Carey, Fausnaugh, Gaudi,
  Henderson, Kerr, Pogge, Varricatt, Wibking, Yee, Zhu, Poleski, Pawlak,
  Szyma{\'{n}}ski, Skowron, Mr{\'{o}}z, Koz{\l}owski, Wyrzykowski,
  Pietrukowicz, Pietrzy{\'{n}}ski, Soszy{\'{n}}ski, Ulaczyk, Choi, Park, Jung,
  Shin, Albrow, Park, Kim, Lee, Cha, Kim, Lee, Maoz, Kaspi, Street, Tsapras,
  Bachelet, Dominik, Bramich, Horne, Snodgrass, Steele, Menzies, Jaimes,
  Wambsganss, Schmidt, Cassan, Ranc, Mao, Dong, D'Ago, Scarpetta, Verma,
  J{\o}rgensen, Kerins, \& Skottfelt}]{Shvartzvald2015-spitzer}
Shvartzvald, Y., Udalski, A., Gould, A., {et~al.} 2015, ApJ, 814, 111

\bibitem[{Smullen {et~al.}(2016)Smullen, Kratter, \& Shannon}]{Smullen2016}
Smullen, R.~A., Kratter, K.~M., \& Shannon, A. 2016, eprint arXiv:1604.03121

\bibitem[{Spergel {et~al.}(2015)Spergel, Gehrels, Baltay, Bennett,
  Breckinridge, Donahue, Dressler, Gaudi, Greene, Guyon, Hirata, Kalirai,
  Kasdin, Macintosh, Moos, Perlmutter, Postman, Rauscher, Rhodes, Wang,
  Weinberg, Benford, Hudson, Jeong, Mellier, Traub, Yamada, Capak, Colbert,
  Masters, Penny, Savransky, Stern, Zimmerman, Barry, Bartusek, Carpenter,
  Cheng, Content, Dekens, Demers, Grady, Jackson, Kuan, Kruk, Melton, Nemati,
  Parvin, Poberezhskiy, Peddie, Ruffa, Wallace, Whipple, Wollack, Zhao,
  Spergel, Gehrels, Baltay, Bennett, Breckinridge, Donahue, Dressler, Gaudi,
  Greene, Guyon, Hirata, Kalirai, Kasdin, Macintosh, Moos, Perlmutter, Postman,
  Rauscher, Rhodes, Wang, Weinberg, Benford, Hudson, Jeong, Mellier, Traub,
  Yamada, Capak, Colbert, Masters, Penny, Savransky, Stern, Zimmerman, Barry,
  Bartusek, Carpenter, Cheng, Content, Dekens, Demers, Grady, Jackson, Kuan,
  Kruk, Melton, Nemati, Parvin, Poberezhskiy, Peddie, Ruffa, Wallace, Whipple,
  Wollack, \& Zhao}]{Spergel2015}
Spergel, D., Gehrels, N., Baltay, C., {et~al.} 2015, eprint arXiv:1503.03757

\bibitem[{Standish \& Williams(2013)}]{Standish2013}
Standish, E.~M., \& Williams, J.~G. 2013, in Explanatory Supplement to the
  Astronomical Almanac, 3rd edn., ed. S.~Urban \& P.~K. Seidelmann (University
  Science Books), 305--346

\bibitem[{Street {et~al.}(2016)Street, Udalski, Novati, Hundertmark, Zhu,
  Gould, Yee, Tsapras, Bennett, J{\o}rgensen, Dominik, Andersen, Bachelet,
  Bozza, Bramich, Burgdorf, Cassan, Ciceri, D'Ago, Dong, Evans, Gu, Harkonnen,
  Hinse, Horne, Jaimes, Kains, Kerins, Korhonen, Kuffmeier, Mancini, Menzies,
  Mao, Peixinho, Popovas, Rabus, Rahvar, Ranc, Rasmussen, Scarpetta, Schmidt,
  Skottfelt, Snodgrass, Southworth, Steele, Surdej, Unda-Sanzana, Verma, von
  Essen, Wambsganss, Wang, Wertz, Poleski, Pawlak, Szyma{\'{n}}ski, Skowron,
  Mr{\'{o}}z, Koz{\l}owski, Wyrzykowski, Pietrukowicz, Pietrzy{\'{n}}ski,
  Soszy{\'{n}}ski, Ulaczyk, Beichman, Bryden, Carey, Gaudi, Henderson, Pogge,
  Shvartzvald, Abe, Asakura, Bhattacharya, Bond, Donachie, Freeman, Fukui,
  Hirao, Inayama, Itow, Koshimoto, Li, Ling, Masuda, Matsubara, Muraki,
  Nagakane, Nishioka, Ohnishi, Oyokawa, Rattenbury, Saito, Sharan, Sullivan,
  Sumi, Suzuki, Tristram, Wakiyama, Yonehara, Han, Choi, Park, Jung, \&
  Shin}]{Street2016}
Street, R.~A., Udalski, A., Novati, S.~C., {et~al.} 2016, ApJ, 819, 93

\bibitem[{Sumi \& Penny(2016)}]{Sumi2016}
Sumi, T., \& Penny, M.~T. 2016, eprint arXiv:1603.05797

\bibitem[{Sumi {et~al.}(2011)Sumi, Kamiya, Bennett, Bond, Abe, Botzler, Fukui,
  Furusawa, Hearnshaw, Itow, Kilmartin, Korpela, Lin, Ling, Masuda, Matsubara,
  Miyake, Motomura, Muraki, Nagaya, Nakamura, Ohnishi, Okumura, Perrott,
  Rattenbury, Saito, Sako, Sullivan, Sweatman, Tristram, Udalski,
  Szyma{\'{n}}ski, Kubiak, Pietrzy{\'{n}}ski, Poleski, Soszy{\'{n}}ski,
  Wyrzykowski, Ulaczyk, \& {Microlensing Observations in Astrophysics (MOA)
  Collaboration}}]{Sumi2011}
Sumi, T., Kamiya, K., Bennett, D.~P., {et~al.} 2011, Natur, 473, 349

\bibitem[{Sumi {et~al.}(2013)Sumi, Bennett, Bond, Abe, Botzler, Fukui,
  Furusawa, Itow, Ling, Masuda, Matsubara, Muraki, Ohnishi, Rattenbury, Saito,
  Sullivan, Suzuki, Sweatman, Tristram, Wada, \& Yock}]{Sumi2013}
Sumi, T., Bennett, D.~P., Bond, I.~A., {et~al.} 2013, ApJ, 778, 150

\bibitem[{Sutherland \& Fabrycky(2016)}]{Sutherland2016}
Sutherland, A.~P., \& Fabrycky, D.~C. 2016, ApJ, 818, 6

\bibitem[{Udalski {et~al.}(2015{\natexlab{a}})Udalski, Szyma{\'{n}}ski, \&
  Szyma{\'{n}}ski}]{Udalski2015-ogleiv}
Udalski, A., Szyma{\'{n}}ski, M.~K., \& Szyma{\'{n}}ski, G. 2015{\natexlab{a}},
  AcA, 65, 1

\bibitem[{Udalski {et~al.}(2015{\natexlab{b}})Udalski, Yee, Gould, Carey, Zhu,
  Skowron, Koz{\l}owski, Poleski, Pietrukowicz, Pietrzy{\'{n}}ski,
  Szyma{\'{n}}ski, Mr{\'{o}}z, Soszy{\'{n}}ski, Ulaczyk, Wyrzykowski, Han,
  {Calchi Novati}, \& Pogge}]{Udalski2015-spitzer}
Udalski, A., Yee, J.~C., Gould, A., {et~al.} 2015{\natexlab{b}}, ApJ, 799, 237

\bibitem[{{Van Cleve} \& Caldwell(2009)}]{vanCleve2009}
{Van Cleve}, J.~E., \& Caldwell, D.~A. 2009, {Kepler Instrument Handbook}

\bibitem[{Veras {et~al.}(2016)Veras, Mustill, G{\"{a}}nsicke, Redfield,
  Georgakarakos, Bowler, \& Lloyd}]{Veras2016}
Veras, D., Mustill, A.~J., G{\"{a}}nsicke, B.~T., {et~al.} 2016, MNRAS, stw476

\bibitem[{Veras \& Raymond(2012)}]{Veras2012a}
Veras, D., \& Raymond, S.~N. 2012, Monthly Notices of the Royal Astronomical
  Society: Letters, 421, L117

\bibitem[{Witt \& Mao(1994)}]{Witt1994}
Witt, H.~J., \& Mao, S. 1994, ApJ, 430, 505

\bibitem[{Wyrzykowski {et~al.}(2006)Wyrzykowski, Udalski, Mao, Kubiak,
  Szymanski, Pietrzynski, Soszynski, \& Szewczyk}]{Wyrzykowski2006}
Wyrzykowski, Â., Udalski, Â., Mao, Â., {et~al.} 2006, AcA

\bibitem[{Yee {et~al.}(2015)Yee, Udalski, Novati, Gould, Carey, Poleski, Gaudi,
  Pogge, Skowron, Koz{\l}owski, Mr{\'{o}}z, Pietrukowicz, Pietrzy{\'{n}}ski,
  Szyma{\'{n}}ski, Soszy{\'{n}}ski, Ulaczyk, \& Wyrzykowski}]{Yee2015}
Yee, J.~C., Udalski, A., Novati, S.~C., {et~al.} 2015, ApJ, 802, 76

\bibitem[{Zhu {et~al.}(2015{\natexlab{a}})Zhu, {Calchi Novati}, Gould, Udalski,
  Han, Shvartzvald, Ranc, Jorgensen, Poleski, Bozza, Beichman, Bryden, Carey,
  Gaudi, Henderson, Pogge, Porritt, Wibking, Yee, Pawlak, Szymanski, Skowron,
  Mroz, Kozlowski, Wyrzykowski, Pietrukowicz, Pietrzynski, Soszynski, Ulaczyk,
  Choi, Park, Jung, Shin, Albrow, Park, Kim, Lee, Kim, Lee, Friedmann, Kaspi,
  Maoz, Hundertmark, Street, Tsapras, Bramich, Cassan, Dominik, Bachelet, Dong,
  {Figuera Jaimes}, Horne, Mao, Menzies, Schmidt, Snodgrass, Steele,
  Wambsganss, Skottfelt, Andersen, Burgdorf, Ciceri, D'Ago, Evans, Gu, Hinse,
  Kerins, Korhonen, Kuffmeier, Mancini, Peixinho, Popovas, Rabus, Rahvar,
  Rasmussen, Scarpetta, Southworth, Surdej, von Essen, Wang, \&
  Wertz}]{Zhu2015-isolated}
Zhu, W., {Calchi Novati}, S., Gould, A., {et~al.} 2015{\natexlab{a}}, eprint
  arXiv:1510.02097

\bibitem[{Zhu {et~al.}(2015{\natexlab{b}})Zhu, Udalski, Gould, Dominik, Bozza,
  Han, Yee, Novati, Beichman, Carey, Poleski, Skowron, Koz{\l}owski,
  Mr{\'{o}}z, Pietrukowicz, Pietrzy{\'{n}}ski, Szyma{\'{n}}ski,
  Soszy{\'{n}}ski, Ulaczyk, Wyrzykowski, Gaudi, Pogge, DePoy, Jung, Choi,
  Hwang, Shin, Park, \& Jeong}]{Zhu2015-binary}
Zhu, W., Udalski, A., Gould, A., {et~al.} 2015{\natexlab{b}}, ApJ, 805, 8

\end{thebibliography}

\end{document}